\documentclass[aps,prb,twocolumn,groupedaddress,showpacs,floatfix]{revtex4}
\usepackage{graphicx}
\usepackage{epsfig}
\begin{document}

%Title of paper
\title{Quantitative Determination of the Adiabatic Condition\\Using Force-Detected
 Nuclear Magnetic Resonance}

% repeat the \author .. \affiliation  etc. as needed
% \email, \thanks, \homepage, \altaffiliation all apply to the current
% author. Explanatory text should go in the []'s, actual e-mail
% address or url should go in the {}'s for \email and \homepage.
% Please use the appropriate macro foreach each type of information

% \affiliation command applies to all authors since the last
% \affiliation command. The \affiliation command should follow the
% other information
% \affiliation can be followed by \email, \homepage, \thanks as well.
\author{Casey W. Miller}
%\email{cmiller@physics.utexas.edu}
\altaffiliation{Present address: Department of Physics, University
of California, San Diego, 9500 Gilman Drive, La Jolla, CA 92093,
USA}
%\affiliation{University of Texas at Austin}
\author{John T. Markert}
\affiliation{Department of Physics, University of Texas at Austin,
1 University Station, Austin, Texas 78712, USA}
\author{Submitted to Physical Review B, June 9, 2005, revised September 20, 2005, accepted October 13, 2005.\\
\large{Phys. Rev. B \textbf{72}, 224402 (2005)}}

\begin{abstract}
The adiabatic condition governing cyclic adiabatic inversion of
proton spins in a micron-sized ammonium chloride crystal was
studied using room temperature nuclear magnetic resonance force
microscopy. A systematic degradation of signal-to-noise was
observed as the adiabatic condition became violated. A theory of
adiabatic following applicable to cyclic adiabatic inversion is
reviewed and implemented to quantitatively determine an
adiabaticity threshold $(\gamma H_1)^2/(\omega_{osc}\Omega)~=~6.0$
from our experimental results.
\end{abstract}

% insert suggested PACS numbers in braces on next line
% insert suggested keywords - APS authors don't need to do this
%\keywords{}

%\maketitle must follow title, authors, abstract, \pacs, and \keywords
\maketitle

% body of paper here - Use proper section commands
% References should be done using the \cite, \ref, and \label commands

\section{Introduction}

The theory of magnetic resonance force microscopy (MRFM) was first
presented by John Sidles in 1991 \cite{Sidles0}. A recent
experiment succeeded in the first MRFM-based detection of a single
electron spin \cite{SingleSpin}. The basic idea of MRFM is that if
the magnetic moment, $\bf m$, of a sample is modulated in time in
the presence of a magnetic field gradient,
{\boldmath$\nabla\,$}${\bf B}$, then there will be an oscillatory
force coupling the two given by ${\bf F}(t) = {\bf
m}(t)\,${\boldmath$\cdot\,\nabla$}${\bf B}$.
\begin{figure}
\begin{center}
\epsfig{file=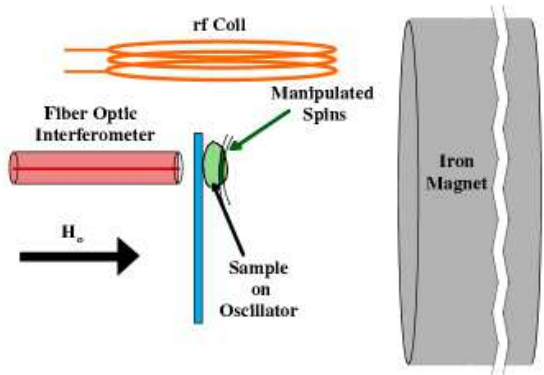,width=3.0in} %
%\epsffile[131 107 498 475]{ad_mvsomega.eps}
     \caption{\small{(Color online) Experimental setup for the
     sample-on-oscillator configuration used in this study.}}
    \label{sampleon}
\end{center}
\end{figure}
\noindent In practice, as illustrated in Fig.~\ref{sampleon} for
MRFM using nuclear spins (NMRFM), the sample is mounted on a low
spring constant mechanical oscillator, the field gradient is
supplied by a permanent magnet, and the nuclear magnetization is
manipulated using radio frequency ($rf$) magnetic fields;
alternatively, one places a small magnet on the oscillator in order
to scan arbitrary samples\cite{MDC_Fab,Choi,Belgium,Hammel}. The
deflection of the oscillator due to the force ${\bf F}(t)$ is
detected with a fiber optic interferometer. If the sample's
magnetization varies in time with a frequency equal to the resonance
frequency of the oscillator, then the deflection amplitude is
increased by the quality factor, Q, of the oscillator. The minimum
detectable force of an NMRFM experiment is limited by the thermal
noise of the mechanical oscillator, given by $
F_{min}~=~\sqrt{4k_{osc}\,k_BT\,\Delta\nu/\left(\omega_{osc}Q\right)}$,
where $k_{osc}$ and $\omega_{osc}$ are the spring constant and
resonance frequency of the mechanical oscillator, $k_B$ is
Boltzmann's constant, $T$ is the temperature, and $\Delta\nu$ is the
equivalent noise bandwidth of the measurement\cite{SidlesRMP}.

The relaxation of spins during measurement is a detriment to
signal strength, and is a subject of ongoing research
\cite{Mozy,Berman,Sutter,Alloul1,Rugar1,Rugar2,Rugar3}.  The
regimes of spin manipulation can be classified by the relative
magnitudes of the spin relaxation rate and the resonance frequency
of the mechanical oscillator.  Cyclic adiabatic inversion,
interrupted cyclic adiabatic inversion, and cyclic saturation are
applied when the relaxation rate is much less than, of the same
order, or much greater than the resonance frequency of the
mechanical oscillator, respectively. The focus of this study is on
cyclic adiabatic inversion and the level of adiabaticity required
to manipulate the maximum number of spins for a given set of
experimental parameters when relaxation is negligible.

\section{Cyclic Adiabatic Inversion}

Cyclic adiabatic inversion is superficially identical to the
conventional adiabatic rapid passage technique \cite{Abragam}.
However, instead of one sweeping pass, cyclic adiabatic inversion
repeats many adiabatic rapid passages so as to make the ${\hat z}$
component of the sample magnetization, $M_z$, periodic. In samples
with relaxation rates much less than the oscillator resonance
frequency, spins will lock to the effective field throughout a
truly adiabatic cyclic inversion. The slowly varying effective
magnetic field essentially looks static to such spins, and they
readily precess around and follow it as it changes in time.

The description of cyclic adiabatic inversion is mathematically
straightforward. The effective field in the rotating frame
(designated by primed unit vectors), ${\bf H}_{eff}$, can be
generically written as $\left(H_o - \omega_{rf}/\gamma\right){\hat
z}~+~H_1{\hat x'}$, where $H_o$ is the total polarizing field,
$\gamma$ is the gyromagnetic ratio of the sample spins, and
$\omega_{rf}$ and $H_1$ are the angular frequency and magnitude of
the rotating $rf$ field, respectively. Cyclic adiabatic inversion is
achieved by frequency modulation (fm) of the $rf$ field. The carrier
frequency is the Larmor frequency of the spins $\omega_o =
H_o/\gamma$, the modulation frequency is the resonance frequency of
the mechanical oscillator $\omega_{osc}$, and the fm amplitude is
$\Omega$. The ${\hat z}$ component of the effective field can thus
be written ${\bf H}_{eff}${\boldmath$\cdot$}${\hat z} = H_o -
(\omega_o+\Omega\sin(\omega_{osc}t))/\gamma$. The ${\hat z}$
component of the magnetization of the sample has the same time
dependence since the spins are locked to the oscillating effective
field. The oscillating sample magnetization in the presence of a
field gradient thus results in a time varying force that resonantly
excites the mechanical oscillator.

In its simplest form, the adiabatic condition governing cyclic
adiabatic inversion says that the effective Larmor frequency of
the spins must be much greater than the rate of change of the
effective field. Translating this into a specific mathematical
statement depends on the particular inversion scheme, but in
general can be written as $\left(\gamma H_{eff}\right)_{min} \gg
\left(d\phi/dt\right)_{max}$, where $\left(\gamma
H_{eff}\right)_{min}$ is the minimum Larmor frequency of the
spins, $\left(d\phi/dt\right)_{max}$ is the maximum angular
frequency of the effective field in the rotating frame, and $\phi$
is defined by $\tan{\phi}~=~({\bf
H}_{eff}${\boldmath$\cdot$}${\hat z})/({\bf
H}_{eff}${\boldmath$\cdot$}${\hat x'})$. In our case, we have
${\bf H}_{eff}~=~\left(H_1, 0,
(\Omega/\gamma)\sin(\omega_{osc}t)\right)$. Taking $\left(\gamma
H_{eff}\right)_{min}~=~\gamma H_1$, and evaluating the time
derivative of $\phi$, leads to the statement of the adiabatic
condition specific to sinusoidal cyclic adiabatic inversion:
\begin{eqnarray}\label{cadicondition}
\frac{(\gamma H_1)^2}{\omega_{osc}\Omega} \gg 1.
\end{eqnarray}

\noindent Written in this manner, the adiabatic condition compares
experimental parameters to unity, with a large number implying the
adiabatic condition is well met.

Equation \ref{cadicondition} implies the importance of the three
major experimental parameters $\omega_{osc}$, $H_1$, and $\Omega$.
Of these, $\Omega$ is the most experimentally flexible;
$\omega_{osc}$ is fixed by the structure of the oscillator, and in
this experiment a spurious signal artifact associated with the
$rf$ limits $H_1$ to about 7~G. In contrast, $\Omega$ can be
changed over two decades.

There is a limit to the magnitude of the magnetization we can
manipulate for a given set of parameters, even if we assume the
adiabatic condition is well met and do not consider relaxation.
The time varying $\hat{z}$ component of the magnetization can be
written as
\begin{eqnarray}\label{mdcMz}
M_z(t) =
M_o\frac{\frac{\Omega}{\gamma}\sin\omega_{osc}t}{\sqrt{(\frac{\Omega}{\gamma}\sin
\omega_{osc}t)^2+H_1^2}}\nonumber,
\end{eqnarray}\noindent where $M_o$ is the equilibrium sample magnetization
following Curie's Law. The maximum manipulable magnetization is
thus
\begin{eqnarray}
\label{myMz} \left(\frac{M_z}{M_o}\right)_{max} =
\frac{1}{\sqrt{1+(\frac{\gamma H_1}{\Omega})^2}}.
\end{eqnarray}
We see that the maximum magnetic moment per unit volume contributing
to a force signal through cyclic adiabatic inversion will always be
less than the equilibrium magnetization. The participating
magnetization should increase with increasing $\Omega$, for all
other parameters constant. However, comparison with the adiabatic
condition shows that increasing $\Omega$ decreases the level of
adiabaticity.  There is clearly a competition between these two that
determines the signal strength. If adiabaticity is achieved, the
signal should be consistent with Eq.~\ref{myMz}.  Violations of the
adiabatic condition would result in deviations from this behavior.

\section{Experimental Details}
This proton NMRFM experiment was performed in a uniform 8.073~T
external magnetic field in the sample-on-oscillator configuration
(Fig.~\ref{sampleon}). The experiment was performed at room
temperature in a vacuum of 15~mTorr. An ammonium chloride (NH$_4$Cl)
crystal was mounted onto the head of a double torsional oscillator
using a thin glass fiber and N-grease. The sample was a flat
cylinder roughly 10~$\mu$m thick and 25~$\mu$m in diameter. This
salt was chosen for its abundance of protons ($6.9 \times
10^{22}~^1$H/cm$^3$) and its long room temperature spin-lattice
relaxation time (\textit{T}$_1 \sim$ 3 s). Note that in the short
correlation time regime $T_1~\sim~T_{1\rho}$.

The mechanical oscillator was a single-crystal silicon double
torsional oscillator \cite{mmm8,mmm9,mmmpap}. The first cantilever
mode of the oscillator was used, which had a resonance frequency
of 7~kHz, a pressure-limited quality factor of 900, and an
estimated spring constant of 10$^{-2}$~N/m.  The minimum
detectable force at room temperature using a 2.5 Hz measurement
bandwidth was thus $3.3 \times 10^{-15}$~N.

The permanent magnet that provided the field gradient was an iron
cylinder that was molded in a zirconium gettered, argon atmosphere
arc furnace. The cylinder was 44.0~mm long, and had a radius of
0.76~mm. Modelling was used to estimate the axial field, ${\bf
B}_z$, and field gradient, $\nabla_z{\rm B}$. A reasonable
estimate of the resonance slice thickness (i.e., resolution)
is\cite{VeemanNMRS}
\begin{eqnarray}
\label{slice} \Delta z =\frac{2\Omega}{\gamma \nabla_z{\rm B}}.
\end{eqnarray}
\noindent  For the purpose of this experiment, however, spacial
resolution of the sample was not a concern. Accordingly, we chose
to operate at a comfortable magnet-to-sample distance of 1 mm,
where $\nabla_z{\rm B}$ was 310~T/m.  This field gradient and our
fm amplitude range of $\Omega /2\pi~\in~[5, 40~$kHz] corresponded
to resonance slice thicknesses $\Delta z~\in~[0.8, 6.0~\mu$m]. The
field ${\bf B}_z$ at 1~mm was 0.191~T, resulting in a total
polarizing field of 8.264~T, and a proton Larmor frequency of
351.8~MHz. Using Curie's Law to calculate the equilibrium
magnetization of the sample, and taking the volume of spins
contributing to the signal to be $a\Delta z$, where $a$ is the
cross sectional area of the sample, the theoretical force for
$\Omega/2\pi$ = 40~kHz was $5.7\times~10^{-14}$~N, leading to a
theoretical signal-to-noise ratio (SNR) of 17; the theoretical SNR
for $\Omega/2\pi$ = 5~kHz was 2.

For each value of $\Omega /2\pi$, with a constant $H_1$ of 7~G, the
iron magnet was moved in 3~$\mu$m steps so that the resonance slice
scanned through the sample. We performed and averaged 5-8 cyclic
adiabatic inversion sequences at each position. The signal from each
position was determined using the \textit{rms} value of a 100~ms
long section of the time series, starting $\sim$30~ms after the
oscillator reached its driven equilibrium amplitude. Due to the long
relaxation time of proton spins in NH$_4$Cl, the temporal response
of the signal showed no degradation over the investigated time
scale.  The baseline was subtracted from the signal at each position
to determine the NMR-induced oscillator amplitude. The NMR origin of
the signal was verified by observing the signal shift $100~\mu$m
when the $rf$ carrier frequency was increased by 1~MHz.

\begin{figure}
\begin{center}
\epsfig{file=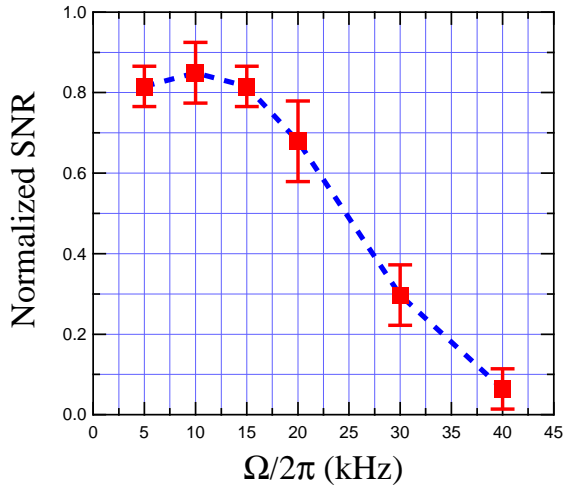,width=3.3in} %
    \caption{\small{(Color online) Normalized SNR
    as a function of the frequency modulation amplitude.  The
    observed decrease occurs as the adiabatic condition is violated.}}
    \label{omegadata}
\end{center}
\end{figure}

Figure~\ref{omegadata} presents normalized SNR (nSNR) data as a
function of the frequency modulation amplitude taken with the
resonance slice in the center of the sample. We define nSNR as the
experimental SNR normalized by the theoretical SNR for each
$\Omega$. Using nSNR implicitly accounts for signal changes due to
the $\Omega$-dependence of both the maximum manipulable
magnetization (Eq.~\ref{myMz}) and the resonance slice thickness
(Eq.~\ref{slice}). Thus, when the measured signal is in accord
with the expected signal, nSNR should be independent of the
experimental parameters that comprise the adiabaticity factor. The
data of Fig.~\ref{omegadata} show such behavior for $\Omega/2\pi$
 of $15~{\rm kHz}$ and below.  In this region, the maximum number of
spins per unit volume were engaged in cyclic adiabatic inversion.
In contrast, the adiabatic condition became violated as
$\Omega/2\pi$ was increased above $15~{\rm kHz}$. Here, the
effective field moved too rapidly for the proton spins to follow,
resulting in a degradation of nSNR. Similar behavior can be
inferred from $^{19}$F data in CaF$_2$ from a previous
study\cite{RugarCAdI}. A systematic error in calculating the
theoretical signal or theoretical sensitivity is most likely
responsible for the nSNR not reaching unity; this error has no
$\Omega$ dependence.

\section{Theoretical Formalism}
Here we describe a theory set forth by Sawicki and Eberly in the
context of quantum optics \cite{Eberly}, but now discussed in the
framework of nuclear magnetism and cyclic adiabatic inversion. An
essential assumption for the reasonable application of this theory
is that the relaxation of locked spins is negligible. This
requirement was met in our case as the relevant relaxation times of
the proton spins were three orders of magnitude larger than the time
for a single inversion $\pi/\omega_{osc}$, and more than one order
larger than the cyclic adiabatic inversion time series. We propose
that the following theory may be applied to multiple inversions
without loss of generality when relaxation phenomena may be ignored.

Taking the magnetization in the rotating frame at equilibrium to
be $\bf{M}$$= (0,0,1)$, the time evolution of the Bloch-vector
$\bf{M}$ is given by
\begin{eqnarray}\label{eqn2}
\frac{d\bf{M}}{dt}=\gamma{\bf H}_{eff}\times {\bf M},
\end{eqnarray}
\noindent where $\bf{M}$ rotates about the effective field. As a
first approximation to cyclic adiabatic inversion, assume the
effective field in the rotating frame is uniformly rotating with
constant angular velocity $A$ in the $\hat{x}'$-$\hat{z}$ plane
about the vector ${\bf A} = A\hat{y}'$. The time evolution of
$\gamma {{\bf H}_{eff}}$ is then
\begin{eqnarray}\label{eqn4}
\frac{d (\gamma{{\bf H}_{eff}})}{dt}={\bf A}\times \gamma{{\bf
H}_{eff}}.
\end{eqnarray}
\noindent The effective field then becomes $\gamma{{\bf
H}_{eff}}=(\pm\gamma H_{eff}\sin{At},0,\pm\gamma
H_{eff}\cos{At})$. The spin-locking solution that satisfies both
Eq.~\ref{eqn2} and Eq.~\ref{eqn4} is ${\bf M}=\pm\left(\gamma{{\bf
H}_{eff}}(t)-{\bf A}\right)\xi$. The constant $\xi$ has units of
time and is under the constraint $\bf{M}\cdot\bf{M}~=$~1, which is
to say $(\gamma H_{eff}^2+A^2)\xi^2 = 1$, where $A$ is the angular
velocity of $\gamma {{\bf H}_{eff}}$ about $\bf{A}$. The exact
solution for the Bloch vector in the rotating frame is
%\begin{widetext}
\begin{eqnarray}\label{BlochVector}
%{\bf M} = \left(AT\sin{(t/T)}\cos{(At)}-(1+A^2T^2(\cos{(t/T)}-1)\sin{(At)})\right)\hat{x}'\nonumber\\
%+~~\left(\gamma H_{eff} A T^2(\cos{(t/T)}-1)\right)\hat{y}'\nonumber\\
%+~~\left(-AT\sin{(t/T)}\sin{(At)}-(1+A^2T^2(\cos{(t/T)}-1))\cos{(At)}\right)\hat{z}.\nonumber
M_{x'} = A\xi\sin{(t/\xi)}\cos{(At)}\nonumber\\-(1+A^2\xi^2(\cos{(t/\xi)}-1)\sin{(At)}),\nonumber\\
M_{y'} = \gamma H_{eff} A \xi^2(\cos{(t/\xi)}-1)\nonumber,\\
M_{z} =
-A\xi\sin{(t/\xi)}\sin{(At)}\nonumber\\-(1+A^2\xi^2(\cos{(t/\xi)}-1))\cos{(At)}.\nonumber
\end{eqnarray}
%\end{widetext}

The object of interest for adiabatic inversion is the probability
of spins following $\gamma {{\bf H}_{eff}}$ through a rotation of
$\pi$ radians, $P_{\pi}$. Assuming no spin relaxation, a trivial
extension of the work of Sawicki and Eberly finds $P_{\pi}$ for
the inversion time $t=\pi/A$ to be
\begin{eqnarray}\label{pescx}
P_{\pi}=1-\frac{\Lambda^2}{1+\Lambda^2}\sin
^2\left(\frac{\pi}{2}\frac{\sqrt{1+\Lambda^2}}{\Lambda}\right),
\end{eqnarray}
\noindent where we have introduced a diabaticity parameter
$\Lambda\equiv A/\gamma H_{eff}$.  This parameter compares the
angular velocity of the rotating effective field to the angular
velocity of the spins about the effective field.  Thus,
$P_{\pi}\rightarrow 1$ in the adiabatic limit of $\gamma
H_{eff}\gg A$, and $P_{\pi}\rightarrow 0$ in the diabatic limit of
$A\gg\gamma H_{eff}$.\\
\begin{figure}
\begin{center}
\epsfig{file=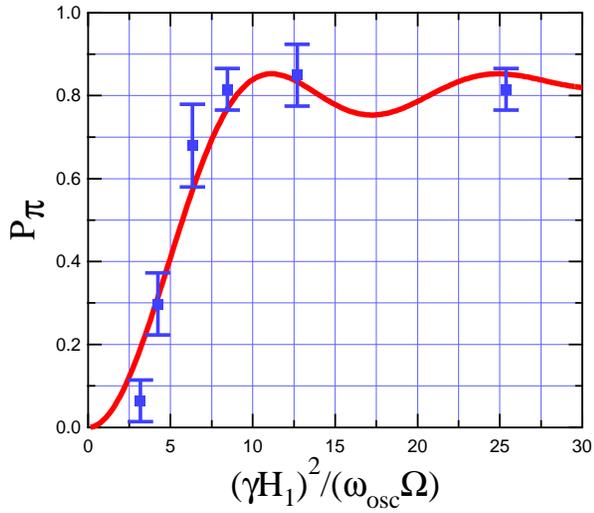,width=3.3 in} %
%\epsffile[136 137 505 502]{ad_followzoom.eps}
    \caption{\small{(Color online) Fit of experimental data using
    the adiabatic following theory of Eq.~\ref{pescx}.}}
    \label{fit}
\end{center}
\end{figure}
\section{Discussion}
To compare experimental cyclic adiabatic inversion data with
Eq.~\ref{pescx}, we make the identification $1/\Lambda~=~(\gamma
H_1)^2/(\omega_{osc}\Omega)$. Furthermore, we introduce two fitting
parameters, $s$ and $n$, and make the transformation
$P_{\pi}(\Lambda)\rightarrow sP_{\pi}(n\Lambda)$. Fig.~\ref{fit}
shows a least squares fit of our experimental data to this function,
where the best fit parameters were $s~=~0.85$ and $n~=~6.4$. We
define the adiabatic threshold as the value of $(\gamma
H_1)^2/(\omega_{osc}\Omega)$ for which the best fit comes within
$1/e$ of its maximum.  We thus determine an adiabatic threshold of
6.0 for protons in NH$_4$Cl.

The non-uniform rotation of the experimental effective field makes
the adiabatic condition of Eq.~\ref{cadicondition} a conservative
statement as it considers the \textit{minimum} Larmor frequency
and the \textit{maximum} angular velocity of the effective field.
Experimentally, both of these, and thus the adiabatic factor, are
time dependent. The adiabaticity threshold of 6.0 corresponds to
the worst case scenario, which is experienced on resonance (${\bf
H}_{eff}~=~H_1{\hat x'}$). Further theoretical development is
necessary to define an adiabaticity threshold for non-uniform
rotation.

The discrepancy between the specific way in which spins are inverted
in this theory and our experiment is a minor one. The theory
considers an effective field uniformly rotating in the
$\hat{x}'$-$\hat{z}$ plane. Experimentally, only the ${\hat
z}$-component of ${\bf H}_{eff}$ is time-dependent. In fact, the
theoretical effective field projected onto the line $x' = H_1$ is
exactly the experimental effective field. If there is true adiabatic
following, the spins should follow both descriptions of the
effective field equally well. Further, the experiment is only
sensitive to the $\hat{z}$-component of the magnetization, which is
identical in the theory and the experiment. It is thus quite
reasonable that this theoretical treatment applies to adiabatic
inversion. However, it is curious, and perhaps specific to this type
of spin system, that a theory concerning a single $\pi$ inversion
applies to cyclic inversions. Additional research is needed to
further elucidate this phenomenon, and to
comment on the general applicability of the theory.\\

\section{Conclusion}
We have investigated the level of adiabaticity necessary to perform
room temperature cyclic adiabatic inversion of proton spins in an
ammonium chloride crystal using nuclear magnetic resonance force
microscopy. We observed a systematic degradation of signal-to-noise
as the adiabaticity factor decreased below this level. A theory of
adiabatic following was discussed to describe cyclic adiabatic
inversion in terms of experimentally relevant parameters, and was
utilized to quantitatively determine an adiabaticity threshold
$(\gamma H_1)^2/(\omega_{osc}\Omega)~=~6.0$ from our experimental
results.

\begin{acknowledgments}
\vspace{-0.1in}This work was supported by the Robert A. Welch
Foundation Grant No. F-1191, the Army Research Office Contract No.
DAAD-19-02-C-0064 through Xidex Corporation, and the National
Science Foundation Grant No. DMR-0210383. CWM acknowledges an
Outstanding Dissertation Award from The University of Texas at
Austin Department of Physics.
\end{acknowledgments}

\end{document}